\documentclass[final,5p,times,twocolumn]{elsarticle}

\usepackage{epsfig}

\usepackage{amssymb}

\usepackage{lineno}

\begin{document}

\begin{frontmatter}



\title{Flashing coherently rotating carbon sticks in $^{24}$Mg+$^{24}$Mg collision}


\author[label1,label2]{M. H. Zhao}
\author[label3]{S. Kun\corref{cor1}}
\ead{ksy1956@gmail.com}
\cortext[cor1]{Corresponding author at: 68/15 John Cleland Crescent, Florey ACT 2615, Australia}
\author[label4]{O. Merlo}
\author[label1]{M. R. Huang}
\author[label1,label2]{Y. Li}
\author[label1]{J. S. Wang}

\address[label1]{Institute of Modern Physics, Chinese Academy of Sciences, Lanzhou, 730000, People's Republic of China}
\address[label2]{University of Chinese Academy of Sciences, Beijing 100049, People's Republic of China}
\address[label3]{Canberra, Australia}
\address[label4]{Zurich University of Applied Sciences, Institute of Applied Simulation Gr\"uental, P.O. Box CH-8820 Waedenswil, Schweiz}

\begin{abstract}
We analyze quasi-periodic oscillations in the angle-averaged ($\Delta\theta_{cm}\simeq 90^\circ\pm 25^\circ $) excitation functions for
the $^{24}Mg+^{24}Mg$ elastic-inelastic scattering and $\alpha$-transfer channels on the energy interval $E_{cm}=44.86-47.76$ MeV.
The period of the energy structures, $\simeq$0.81 MeV, is interpreted as inverse half-period ($\simeq 5\times 10^{-21}$ sec.) of coherent rotation of highly
excited short-lived ($\simeq 3.6\times 10^{-21}$ sec.) chain of a length $\simeq 30$ fm. The rotational wave packet coherence survives (i) the energy relaxation (fully mixing ergodic dynamics)
for the strongly overlapping states with fixed total spin and parity values, and (ii) the strong perturbation of the motion upon a change of the total spin.
 The present discussion rises a number of the questions. For example,
 is rotational coherence of large molecules necessarily destroyed in the conventionally statistical limit of
 structureless non-selective continuum (for fixed total spin and parity values) under the conditions of complete intramolecular energy redistribution and vibrational dephasing in the regime of strong ro-vibrational coupling? For the slow cross-symmetry phase relaxation, quantum coherent superpositions of a large number of complex configurations with, {\sl e.g.}, many different total angular momenta produce image of a rotation of macroscopic object with classically fixed (single) total angular momentum.
 Suppose that the quantum coherent superpositions involving a very large number of different good quantum numbers
 play a role, in a hidden form, in a formation of macroscopic world.
 Then why these quantum superpositions are so stable against quick aging/decay of ordered complex structures preventing or slowing down
 tendencies towards uniform occupation of the available phase space as prescribed by the random matrix theory?
 And what kind of complex macroscopic phenomena may reveal traces of partially coherent quantum superpositions involving
  a huge number of quantum-mechanically different integrals of motion behind of what is referred to as conservation laws in classical physics employed for the description of the macroscopic world?


\end{abstract}

\begin{keyword}
coherent rotation \sep phase relaxation \sep $^{24}Mg+^{24}Mg$ collision \sep nuclear molecules \sep quantum macroscopic transition
\end{keyword}

\end{frontmatter}


\label{}
Femtochemistry provides powerful tools in studying molecular structure and dynamics in
chemistry and biology ~\cite{ZewailV1V2}, ~\cite{ZewailNobLect}. Nowadays,
the use of ultrashort femtosecond and attosecond laser pulses is a widely used technique for the real-time monitoring
of chemical reactions.
One of the successes of femtochemistry is its ability for direct
probe of vibrational and rotational wave packets and their coherent evolution as classical-like, essentially due to the quantum interference,
system-specific trajectories of a transitional state of a given chemical reaction. Precondition for survival of the characteristic time and length scales
for the system-specific wave packet dynamics is that the isolated intermediate complex is not in ergodic state.
 This means that phase relaxation
(dephasing) or dynamical decoherence \cite{CasChirI},
responsible for a spread of the wave packets, is relatively slow, not faster than the life time of the intermediate complex.
This regime of incomplete mixing is beyond applicability of random matrix theory \cite{RMW10}
  which addresses a featureless universal behavior of complex systems in the ergodic limit of complete lose of phase memory and uniform occupation of the available phase space (simple pictorial explanation of the domain of applicability of random matrix theory, for not experts in the field, is given in Sect. V of \cite{arXiv13}).

Unfortunately, the femtochemistry experimental methods are inapplicable to study reactions which are not initiated by the laser pulses. In particular, time evolution of heavy ion collisions is not accessible for the direct monitoring. Additional experimental difficulty to measure time scales of reactions involving relatively light heavy ions is their short duration. Namely, time resolution of order of zeptosecond (10$^{-21}$ sec) would be needed for an accurate reconstruction of heavy ion collisional dynamics. Therefore, the only possible way to access time evolution of heavy ion reactions is provided by the detailed energy dependence of the cross sections, {\sl i.e.} excitation functions. In this respect, the most favorable situation is the presence of both fast (direct) and time delayed processes \cite{PRCrapid99}, \cite{PRC01}, \cite{IJMPE03}, \cite{PRC06}. Then, under the condition of a major contribution of direct processes, the energy variations in the excitation functions are mainly given by the interference between the collision amplitudes corresponding to the direct and time delayed processes. Since the direct process amplitude is energy smooth function, this allows to obtain time dependent {\sl amplitude}, ${\cal P}(t,\theta )$ of the time delayed process \cite{PRCrapid99}, \cite{PRC01}, \cite{IJMPE03}, \cite{PRC06}. Clearly, the direct process plays a role of the pump pulse switching on the clock at the initial moment of time \cite{IJMPE03}, \cite{PRC06}. When, for that or another reason, the interference term between the amplitudes of the direct and time delayed processes does not contribute to the cross section, the energy variations in the excitations functions originate from the modulus square of the energy oscillating around zero amplitude, $F(E,\theta )$, corresponding to the time delayed processes. In this situation, an information on the time evolution, though not as complete as in the presence of direct processes, is still encoded in the incident energy dependence of the cross sections. Indeed, we represent
the cross section $\sigma(E,\theta ) = | F(E,\theta ) |^2$ as
\begin{eqnarray}
\sigma(E,\theta)\propto \int_{-\infty}^\infty d\tau\exp(iE\tau /\hbar )<{\cal P}(t+\tau /2,\theta )\nonumber\\
{\cal P}(t-\tau /2,\theta )^\ast>_t
\label{eq1}
\end{eqnarray}
with
\begin{eqnarray}
<{\cal P}(t+\tau /2,\theta ){\cal P}(t-\tau /2,\theta )^\ast>_t=\int_0^\infty dt \nonumber\\
{\cal P}(t+\tau /2,\theta ){\cal P}(t-\tau /2,\theta )^\ast.
\label{eq2}
\end{eqnarray}
In the above relations, $E > 0$ is the total excitation energy, $t$ is the time, $\theta$ is the emission (scattering) angle and
${\cal P}(t,\theta)\propto \int_{E_1}^{E_2} dE\exp(-iEt/\hbar)F(E,\theta )$ with ${\cal P}(t\leq 0,\theta )=0$ has a finite time resolution,
 $\Delta t\simeq 2\pi\hbar /(E_2-E_1)$.

Suppose that a pronounced quasi-periodic component,
with a period of $\delta E$, is present in $\sigma(E,\theta )$. From model independent Eqs. (1) and (2), this implies quasi-periodicity or
recurrences, at least one recurrence, in ${\cal P}(t,\theta )$ with a period of $\simeq 2\pi\hbar/\delta E$.
If, in addition, the quasi-periodic energy components correlate for given well resolved different exit channels, this implies that
 $<{\cal P}(t+\tau /2,\theta ){\cal P}(t-\tau /2,\theta )^\ast>_t$
 is significantly enhanced at approximately the same, for these different correlated channels, moments $\tau\simeq 2M\pi\hbar/\delta E$ with $M=1, 2, ...$, at least with $M=1$.

 Quasi-periodic structures in the excitation functions of heavy ion collisions often reveal presence of non-statistical effects.
 For example, the oscillating energy structures in the $^{24}Mg+^{24}Mg$ elastic-inelastic scattering \cite{PLB83}, \cite{IndiaZPA} were interpreted in terms of the highly excited coherently rotating hyper-deformed intermediate complex \cite{PRL99}. However, the identification of the non-statistical quasi-periodic structures from the analysis of the cross section energy autocorrelation functions in \cite{PRL99} may look questionable for the following reason. In the statistical regime of random phases of the strongly overlapping resonance levels with different total spin values, the energy autocorrelation functions, $C(\varepsilon)$, have a Lorentzian shape \cite{ErMayer66}. This result is obtained
for very long energy interval on which excitation functions are measured. Yet, for the finite data range analysis of the concrete experimental data sets, one encounters deviations from the theoretical prediction, in particular, in the form of fluctuations around the Lorentzians \cite{DallHall66}. Therefore, even though the oscillations in $C(\varepsilon )$'s have approximately the same period for the different exit channels \cite{IndiaZPA}, \cite{PRL99}, which is inconsistent with the statistical interpretation, additional supportive argumentation for the quasi-periodic non-chaotic behavior is highly desirable. The interpretation \cite{PRL99} in terms of the relatively stable coherent rotation, for the slow phase relaxation, was based on the assumption of the highly excited ($\simeq 10-15$ MeV above Yrast line) intermediate complex with strongly overlapping resonances. Therefore, it should be insightful
to also test this interpretation for transfer channels in the $^{24}Mg+^{24}Mg$ collision. Do the energy oscillating structures, with the same quasi-period as that for the  $^{24}Mg(^{24}Mg,^{24}Mg)^{24}Mg$ elastic-inelastic scattering, present in the transfer channels? Affirmative answer to this question would provide additional supportive indication that
both $^{24}Mg(^{24}Mg,^{24}Mg)^{24}Mg$ elastic-inelastic scattering and the transfer reaction channels  originate from the decay of the same highly excited coherently rotating hyper-deformed intermediate complex.

The reasonings presented above  has motivated us to analyze
the angle-averaged, $\Delta\theta_{cm}\simeq 90^\circ\pm 25^\circ $,
excitation functions for the $^{24}Mg+^{24}Mg$
elastic-inelastic scattering and $\alpha$-transfer channels measured on the energy interval $E_{cm}=44.86-47.76$ MeV \cite{PLB87}, which is considerably smaller
than that ($E_{cm}=42-56$ MeV) on which the data were taken in \cite{PLB83}, \cite{IndiaZPA}.
The analysis in \cite{PLB87} confidently confirmed the non-statistical behavior of the energy structures in these processes.
We extend the analysis \cite{PLB87} concentrating on the quantitative interpretation of the pronounced, transparently visible in
Figs. 2 and 3 of \cite{PLB87}, strongly correlated for different channels, quasi-periodic energy structures.
\begin{figure}[htp]
\centering
\setlength{\abovecaptionskip}{1cm}
\includegraphics[width=8.5cm,angle=270]{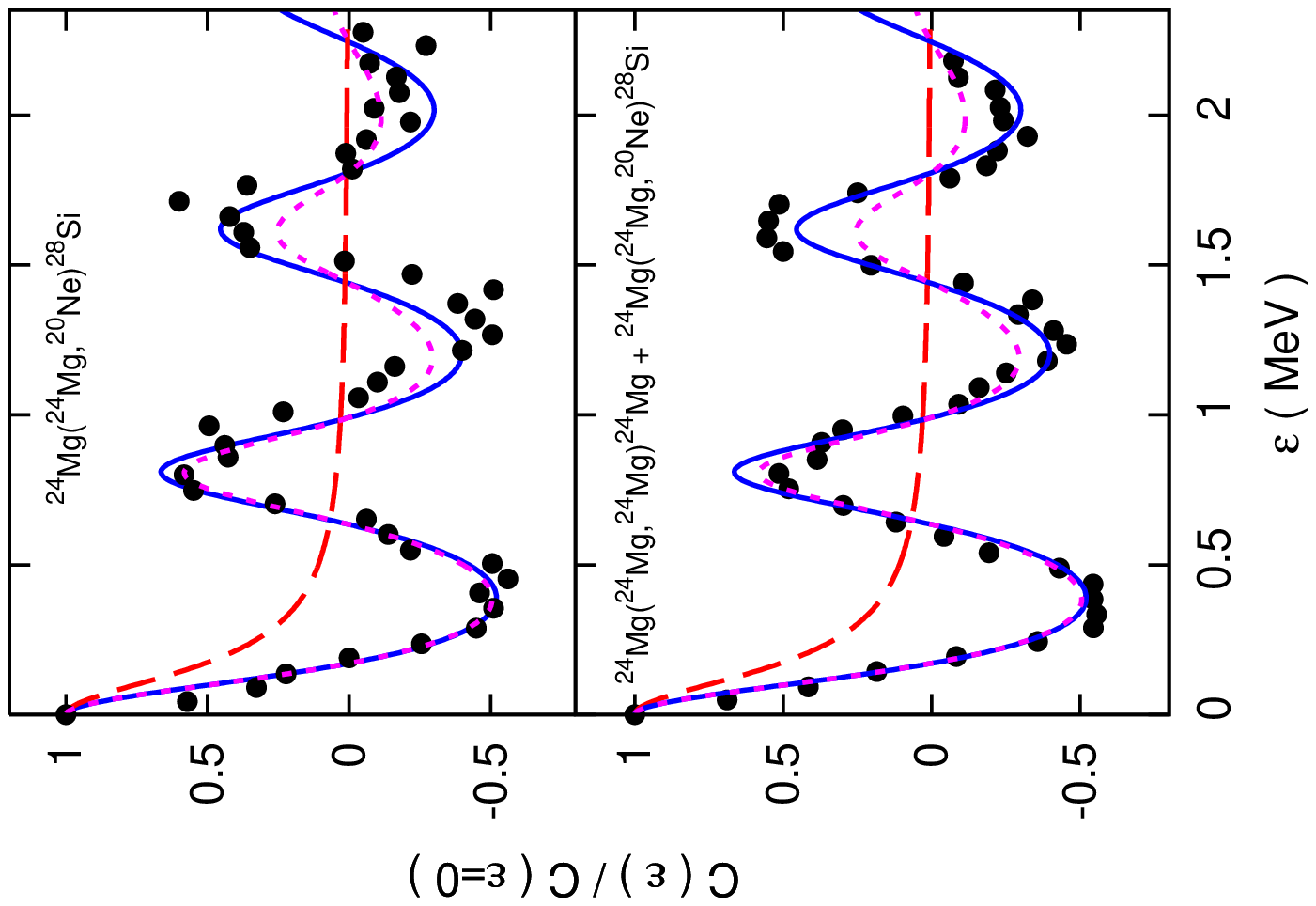}
\caption{\footnotesize
(Color online) The normalized cross section energy autocorrelations functions for the $^{24}Mg(^{24}Mg,^{20}Ne)^{28}Si$ reaction and  $^{24}Mg(^{24}Mg,^{24}Mg)^{24}Mg$ elastic-inelastic scattering. The data (dots) are obtained from the analysis of the summed deviation functions in Fig. 3 of \cite{PLB87}. Solid lines are the fits for the relatively pure angular resolution (though under the neglect of the interference term between the near and far side collision amplitudes)
 obtained with   $\hbar\omega=0.41$ MeV,
$\beta=0.02$ MeV and  $\Gamma=0.2$ MeV. The fit for the poor angular resolution (short-dashed lines) is obtained with   $\hbar\omega=0.404$ MeV,
$\beta=0.0205$ MeV and  $\Gamma=0.17$ MeV. The Lorentzians with $\Gamma=$0.2 MeV (long-dashed lines) are obtained for the short spin off-diagonal phase memory, $\beta >> \Gamma,\hbar\omega$, corresponding to the limit of the random matrix theory \cite{RMW10}.
}			
\label{fig1}
\end{figure}

In Fig. \ref{fig1} we present
the normalized cross section energy autocorrelations functions $C(\varepsilon )$'s constructed from the summed deviation functions in Fig. 3 of \cite{PLB87}. In the upper panel of our Fig. \ref{fig1}, $C(\varepsilon)$ is obtained from the three lowest transitions in the reaction
$^{24}Mg(^{24}Mg,^{20}Ne)^{28}Si$. In the lower panel of our Fig. \ref{fig1}, $C(\varepsilon)$ is obtained from addition of the three lowest well resolved transitions in the reaction $^{24}Mg(^{24}Mg,^{20}Ne)^{28}Si$ and the three lowest well resolved transitions in the  $^{24}Mg(^{24}Mg,^{24}Mg)^{24}Mg$ elastic-inelastic scattering (see Fig. 3 in \cite{PLB87}). The experimental $C(\varepsilon )$'s demonstrate oscillations with the period close to that in Fig. \ref{fig1} of \cite{PRL99}. Therefore, we fit the data with the formula \cite{PRL99}
\begin{eqnarray}
 C(\varepsilon\geq 0)/C(\varepsilon =0)\simeq {\rm Re}\{\exp[i\pi\varepsilon/
(\hbar\omega -i\beta)]
[1- \nonumber\\
\exp[i\pi(\varepsilon + i\Gamma)/(\hbar\omega -i\beta)]]^{-1}\} \nonumber\\
/{\rm Re}\{1/[1-\exp[-\pi\Gamma/(\hbar\omega -i\beta)]]\}.
\label{eq3}
\end{eqnarray}
Here, $\Gamma$ is total decay width of the intermediate complex, $\omega$ is a real part of the angular velocity of the coherent rotation and
$\beta/\hbar$ has a physical meaning of its imaginary part with $\beta$ being the spin off-diagonal phase relaxation width \cite{ZPhysII}.
One observes that $C(\varepsilon )$ (3) oscillates with period $\simeq 2\hbar\omega$. The oscillations are damped for a non-vanishing $\beta$-width.
The fit (solid lines in Fig. 1) is obtained with $\hbar\omega=0.41$ MeV,
$\beta=0.02$ MeV and  $\Gamma=0.2$ MeV. In the limit of short phase memory, $\beta >> \Gamma,\hbar\omega$, $C(\varepsilon )$ has a  Lorentzian shape \cite{ErMayer66}, \cite{RMW10} (long-dashed lines in Fig. 1 with $\Gamma =$0.2 MeV).

A few comments on the derivation and physical interpretation of Eq. (3), in relation to the analyzed data, are in order.

1) The starting point for the derivation of Eq. (3) is a diagonal approximation, with respect to the levels of the intermediate complex, for $S$-matrix elements which can be justified
under the condition of the fast energy redistribution-relaxation within the states with fixed total spin and parity values \cite{arXiv13}.
This diagonal approximation of the Bethe unitary $S$-matrix is valid under the Bethe random sign assumption for partial width amplitudes with given fixed total spin and parity values in a regime of strongly
overlapping resonances providing the partial decay widths are small as comparing with the average level spacing of the highly excited intermediate
  complex, see Section 56D in \cite{Bethe37}. It is important for the present consideration that, unlike the shell-model approach (Section IV(A2) in \cite{RMW10}), the formulation \cite{Bethe37} leads to the Bethe unitary $S$-matrix which is not restricted to the nucleon and $\gamma$-channels only  but is also applicable for binary collisions of composite reaction partners including heavy ions. Indeed, the physical meaningfulness of the diagonal approximation similar to that in \cite{Bethe37} is recognized for heavy ion induced collisions even for a multiple nucleon cascade emission provided the reaction proceeds through an equilibrated intermediate system \cite{Koonin89}. We
  understand that the equilibration in \cite{Koonin89}
  implies (i) energy relaxation and ergodicity within the states with fixed total angular momentum and parity values on each step of the evaporation cascade, and (ii) the conventional assumption of a vanishing
  of the total spin and parity off-diagonal correlations on each step of the evaporation cascade. Yet, while the entrance channel orbital momentum off-diagonal
   interference was neglected making directions along and opposite the incident beam undistinguishable, the strong correlation between different orbital momenta of the evaporated nucleons, on each step of the evaporation cascade, assumed
  in  \cite{Koonin89} (Eqs. (5) and (6) in \cite{Koonin89}) clearly contradicts to the conventional postulate on maximal randomness of the partial width amplitudes (Eq. (2.2) in \cite{FKK80}).
  We mention in passing that the orbital momenta correlations \cite{Koonin89}, if exist in reality, may result in a complete uncertainty of the  standard evaluation of the spin
  dependence of the nuclear level densities
  from angular anisotropy around 90$^\circ$ in the evaporation processes \cite{Str57}, \cite{ErStr58plus}, where such orbital momenta correlations were ignored in a literal consistency
  with the assumption on maximal randomness of the partial width amplitudes.
   As a next step, we extend the argumentation \cite{arXiv13} to justify
 the relations (8.1) and (8.2) in \cite{ZPhysI} between the product of the entrance ($a$) and the exit ($b$) channel partial width amplitudes for different
 $J$-values:
 \begin{equation}
 \gamma_\nu^{Ja}\gamma_\nu^{Jb}=\sum_\mu\gamma_\mu^{Ia}\gamma_\mu^{Ib}Q_{\nu\mu}^{JI} + R_{inc},
 \label{eq4}
 \end{equation}
  where
  \begin{equation}
  Q_{\nu\mu}^{JI }=(1/\pi )
  D\beta|J-I|/[(E_{\nu}^{J}-
  E_{\mu}^I-(J-I)\hbar\omega)^2+\beta^2(J-I)^2]
  \label{eq5}
  \end{equation}
with $E_{\nu}^{J}$ and $E_{\mu}^I$ being resonance energies for the states with total spin values $J$ and $I$, respectively.
The relations (4) hold true independent of whether the energy redistribution process is taken explicitly into account or not \cite{arXiv13}.
We are interested in a regime of strongly overlapping resonances, $\Gamma/D\gg 1$, where both the total decay resonance width $\Gamma$ and the average level spacing $D$ are taken $J$-independent. Under the condition of relatively slow spin off-diagonal phase relaxation, $\beta\ll\Gamma$
(but still $\beta\geq D$), $R_{inc}$ in Eq. (4) can be neglected \cite{ZPhysI} for it produces insignificant contribution into the r.h.s. of Eq. (6).
We obtain
\begin{eqnarray}
S_{ab}^{J}(E)=W[|J-{\bar J}(E)|/d]^{1/2}\sum_\mu
\gamma_{\mu}^{Ia}\gamma_{\mu}^{Ib}/[E- \hbar\omega (J-I)\nonumber\\
-E_\mu^{I}+(i/2)\Gamma+i\beta |J-I|],
\label{eq6}
\end{eqnarray}
where we have omitted energy smooth phase shifts which originate from the potential mean field scattering and direct interaction taking place on the relatively short time intervals of formation and disintegration of the intermediate complex. In case of elastic scattering, $(a=b)$, the products
$\gamma_\nu^{Ja}\gamma_\nu^{Jb}$ and $\gamma_\mu^{Ia}\gamma_\mu^{Ib}$ in Eq. (4) and Eq. (6) should be changed to
$[(\gamma_\nu^{Ja})^2 - \overline{(\gamma_\nu^{Ja})^2}^{\nu}]$ and $[(\gamma_\mu^{Ia})^2 - \overline{(\gamma_\mu^{Ia})^2}^\mu]$, respectively \cite{arXiv13}.
In Eq. (6), $W[|J-{\bar J}(E)|/d]$ is the energy averaged decay probability.
It was taken in a bell shaped form with the maximum at $J={\bar J}(E)$ and with the width $d$ of about an effective range of total spin values coherently excited in the collision process. For the data analyzed in this Letter this range is $\Delta J\simeq 34-38$. In the derivation of Eq. (3), we took ${\bar J}(E)=I+(E-{\bar E})/(\hbar\omega )$, where ${\bar E}$ is the average energy corresponding to $J({\bar E})=I\gg 1$. Under the neglect of the interference term between the near and far side collision amplitudes due to the angle-averaging, such a choice of ${\bar J}(E)$ results in $C(\varepsilon )$ (3) to be approximately independent of $d$. However, for the other choices, {\sl e.g.}  ${\bar J}(E)=I=const$, the additional damping factor, $\simeq W[-|\varepsilon|/(d\hbar\omega)]$, would appear in the r.h.s. of Eq. (3).

2) The data fitted in Fig. 1 were obtained from the angle-averaged excitation functions, $\Delta\theta_{cm}\simeq 90^\circ\pm 25^\circ $. Yet, the formula
(3) was derived for the pure angle-resolution (though under the neglect of the interference term between the near and far side collision amplitudes as well as between direct and time-delayed processes \cite{PRL99}). Generalization for the angle-averaged excitation functions, $\Delta\theta_{cm}\simeq 90^\circ\pm 25^\circ $, yields the additional damping factor in the r.h.s. of Eq. (3). This factor is approximately given by $\sin^2[\varepsilon\Delta\theta/(2^{3/2}\hbar\omega )]/[\varepsilon\Delta\theta/(2^{3/2}\hbar\omega )]^2$ with $\Delta\theta=50^\circ=0.88$. The corresponding fit of the data on the lower panel of Fig. 1 is obtained with $\hbar\omega=0.404$ MeV,
$\beta=0.0205$ MeV and  $\Gamma=0.17$ MeV. Clearly, the additional damping factor, due to the poor angular resolution,
does not affect quasi-period of the oscillations in $C(\varepsilon )$.

3) The time dependent intensity,  $| {\cal P}(t,\theta) |^2$, closely relates to the return probability for a finite time resolution, while modulus square of Fourier component of ${\cal P}(t,\theta)$, {\sl i.e.} the cross section (1), is analogous to the low-resolution version the spectrum \cite{HellerPRL84}
 providing the Bethe $S$-matrix diagonal approximation is applicable (Eq. (260) and Section 56D in \cite{Bethe37}). In our case the finite time resolution is
 $\hbar /\Gamma$, and the spectrum is not resolved due to the strong overlap of the resonance levels, $\Gamma /D\gg 1$.
Clearly, $C(\varepsilon )$ has a physical meaning of the autocorrelation function of the low-resolution version of the spectrum \cite{HellerPRL84}.

In a view of the relatively short energy interval of the measurements \cite{PLB87}, $\Delta E_{cm}=2.9$ MeV, as comparing with $\Gamma$ the standard
evaluation of errors due to the finite data range \cite{DallHall66} would result in the statistical insignificance of the quasi-periodic structures
 in Fig. 1. However, the method of evaluation of the statistical uncertainties \cite{DallHall66} is not applicable in the presence of the $S$-matrix
 spin off-diagonal correlations for $\beta\ll\Gamma$. The essence of the matter can be explained as follows. Consider a given realization of $\sigma^{I}(E)=
|S_{ab}^{I}(E)|^2$, where $S_{ab}^{I}(E)$ is given by Eq. (6), {\sl i.e.} like in \cite{DallHall66}. Then, in a regime of strongly overlapping resonances, the standard
evaluation of errors due to the finite data range \cite{DallHall66} would be certainly applicable and quasi-periodic oscillations in the corresponding $C(\varepsilon )$,
if occur at all,
would not be of the sufficient statistical significance for the short energy interval, $\Delta E_{cm}=2.9$ MeV. Consider next  $\sigma (E)=\sum_{n=-q}^{q}\sigma^{I}(E+2n\hbar\omega )$, where $n$ and $q$ are natural numbers and $n$ in the summation changes with the step of unity.
For $q\gg 1$, $\sigma(E)$ is close to periodic function with the period of $2\hbar\omega$ independent of actual realization of $\sigma^{I}(E)$ the latter being generated like in \cite{DallHall66}.
Clearly, the correspondent $C(\varepsilon\leq r\hbar\omega )$, where $1\ll r\ll q$, will show periodic behavior even if it constructed from $\sigma(E)$ restricted to the
relatively short energy interval $\Delta E\simeq (6-7)\hbar\omega$. For $q=1$ (the three terms in the sum), similar to our case, $\sigma(E)$ will not be exactly periodic but
will still have the strong quasi-periodic component which will show up in the statistically significant way in the corresponding $C(\varepsilon )$ extracted even on the short energy interval $\Delta E\simeq (6-7)\hbar\omega$. The numerical simulations would represent the straightforward way to demonstrate the statistically significant registration of the quasi-periodicity for the short energy intervals. In a view of a simplicity of the simulations these can be  performed during couple of days
by a science or nowadays even, {\sl e.g.}, economy or linguistic undergraduate student. The results could be reported, {\sl e.g.}, during the workshop
``Information and statistics in nuclear experiment and theory ISNET-3"
 at the
ECT$^\ast$ on November 16-20, 2015. In addition to the $^{24}Mg+^{24}Mg$ collisions the simulations may also include
the $^{12}C+^{24}Mg$ elastic and inelastic scattering
which demonstrate strong quasi-periodic structures (Fig. 1 in \cite{PRC01}). This is not a joke for the issue needs clarification because when one of us (S.K.) 15 years ago discussed
 the experimentally observable oscillations in Fig. 1 of \cite{PRC01} with A. Richter (previous Director of the ECT$^\ast$) he was told that these oscillations are due to the finite data range effects and the data are consistent with the random matrix description in terms
of Ericson fluctuations \cite{RMW10}. Then one of the questions to be answered by the student is: How many statistically independent generations,
of the type of \cite{DallHall66}, of the excitation functions
using the standard algorithms of the random matrix theory (the standard statistical model) \cite{RMW10} must be performed in order to reproduce the oscillations in Fig. 1 of \cite{PRC01}? Obviously the student must have an optimistic vision for the future and be brave enough to improve
the world we live in. This is because the opposition is too strong, crossed all possible red lines and, therefore, has nothing to lose \cite{arXiv13}.
For we are speaking about those who demonstrated the inquisitory attitude and are killers of the new ideas which does not leave much choice at this stage. We do not mean here and put aside the Australian way
proved to bring this kind of attitude up to the national operational policy which, bedazzled by the ideas of universality of the random matrix theory, is spiritually oriented on the future with no distinguishability between living human beings and corpses \cite{arXiv13}.

The work under the extended version of this Letter, where the derivation of the cross section energy autocorrelation (3) and some other relevant issues will be addressed, is under way. This will include a discussion of adequacy of the conventional interpretation of the non-statistical structures in terms
of isolated resonances. For example, what is angular dependence of $C(\varepsilon )$, in particular, for the elastic  $^{24}Mg(^{24}Mg,^{24}Mg)^{24}Mg$
scattering in the presence of dominant potential scattering contribution ($\theta\leq 70^\circ$),
 predicted within our treatment in terms of the coherent rotation of the intermediate complex with strongly overlapping resonances as compared with that resulted from the interpretation in terms of isolated resonances as identified in \cite{PRC90}?

Stability of the coherent rotation may be appreciated by noticing that the spreading of the angular orientation during one revolution due to the finite $\beta$-width is $2\pi\beta/\hbar\omega\simeq 15^\circ$. Therefore, it takes about 10 complete revolutions, {\sl i.e.} $\simeq 10^{-19}$ sec., for the complex to loose the coherent nature of its rotation. Energy relaxation time scale for excited nuclei is $\simeq \hbar /\Gamma_{spr}\simeq 10^{-22}$ sec., where $\Gamma_{spr}\simeq 5$ MeV is the spreading width \cite{RMW10}. We observe that the coherent rotation persists for about three orders of magnitude longer than it takes to complete a process of energy equilibration (vibrational relaxation by the molecular physics terminology).

The strong channel correlation between individual well resolved channels in both the  $^{24}Mg(^{24}Mg,^{24}Mg)^{24}Mg$  elastic-inelastic scattering and $^{24}Mg$ $(^{24}Mg,^{20}Ne)^{28}Si$  transfer reaction,
as well as between the channels for these two processes, is transparently visible in Fig. 2 of \cite{PLB87}. Therefore, it is obvious that strongly correlated regular oscillations with the approximately channel-independent quasi-period, $\simeq 0.81$ MeV, are present in all the six channels.
This rules out statistical origin of these strongly correlated quasi-periodic structures and, therefore, their interpretation in terms of Ericson fluctuations \cite{ErMayer66}. This is because Ericson fluctuations, which is a particular case of the Bethe statistical fluctuations (to be reported in the extended version of this Letter), produce uncorrelated, for different channels, irregular energy structures. As a result, the deviations from the Lorentzian must be uncorrelated for different channels instead of demonstrating regular oscillations with the approximately channel-independent quasi-period for all the six channels. The fact that the present analysis of the data on the energy interval $E_{cm}=44.86-47.76$ MeV revealed a value of the quasi-period, $\simeq 0.81$ MeV, very close to that obtained for the much longer energy interval $E_{cm}=42-56$ MeV \cite{PRL99}, supports our interpretation.
A possibility of further test of our interpretation, {\sl e.g.} for the  $^{24}Mg(^{24}Mg,^{24}Mg)^{24}Mg$  elastic scattering, is suggested by the expected dominance of direct processes (potential scattering) for $\theta\leq 70^\circ$ (see Figs. 6 and 7 in \cite{PRC90}).
Then the energy variations in the excitation functions originate mainly from the interference between the collision amplitudes corresponding to the direct and time delayed processes. As a result, the characteristic quasi-periods of leading harmonics in $\sigma (E,\theta)$ become strongly $\theta $-dependent
 \cite{PRC01}, \cite{IJMPE03}, \cite{PRC06}, which can be tested in experiments with pure angular resolution. We mean the numerical experiments against
 the previously measured relevant available data since
 the experimental excavation into the matter as well as the understanding of the underlying phenomena are no longer among interests of the nuclear physics community.

 The numerical experiments are useful to illustrate simple algorithms of transforming/processing the complex quantum information.
 The initial wave packet is given by a coherent superposition of partial waves with different
 orbital momenta describing the colliding objects in their ground states. In a process of the collision the initial quantum superpositions are transformed
 into the coherent superpositions of a large number of strongly overlapping resonance configurations of the intermediate complex with different total spin values. The spin off-diagonal correlations produce the coherent rotation of the intermediate complex with a well defined angular velocity. Yet, the
 intermediate complex represents a quantum superposition of the two objects simultaneously rotating in opposite directions. For these coherently rotating in opposite
 directions alternatives, each of them individually is itself result of the spin off-diagonal quantum interference, produce interference fringes \cite{PRC06}. For small $\beta$-width, the rotating wave packets slowly spread and interference fringes, related to the spin off-diagonal correlations, disappear \cite{PRA07}.
 In this way the complex collision experiments provide a physical visualization of executing the quantum computing algorithms for a simple description of
 the quantum-macroscopic transition, where the convergence to the macroscopic-like dynamics (rotation) on the final stage of the ``calculations" is due
 to partial dynamical decoherence but still images essentially from the quantum
 interference. Each step of the ``quantum computing", presumably run by nature, can be easily mapped on the fundamental double slit experiment
 (since the many slit experiment is a combination of the double slit ones). The double slit
  experiment ``has been designed to contain all of the mystery of quantum mechanics ..." \cite{Feynman65} and, Feynman continues, ``Any other
 situation in quantum mechanics, it turns out, can always be explained by saying, 'You remember the case of the experiment with the two holes? It's the same thing'."  \cite{Feynman65}.
  The
 above simple picture of transformation of the complex quantum information is noticeably supported by the state of art numerical simulations for, {\sl e.g.}, H+D$_2\to$HD+D molecular reaction (Fig. 1 in \cite{JPhChemA06}) even though the calculations \cite{JPhChemA07} rules out a picture of the isolated resonances of the intermediate complex (Fig. 1 in \cite{JPhChemA07}). We mention in passing that the described above algorithms are incompatible with basic ideas of the
 random matrix theory as applied to classically chaotic systems \cite{RMW10}. Therefore it is inappropriate to rely on mentality of indistinguishability and
 selective blindness of the random matrix theory \cite{arXiv13}, which unfortunately found its way beyond the mere academic and education activities, as a point of reference in discussing the above mentioned system specific algorithms.

Angular velocity of coherent rotation, $\omega $, can not be defined within the states with fixed $J$-value. This transparently follows from the
conjecture \cite{arXiv13} which relates $Q_{\nu\mu}^{JI}$ (5) to the spin off-diagonal correlations between squares of the individual resonance wave functions of the intermediate
complex. Indeed, it is obvious that distribution of the spacial density, including the radial extension, is different for the state with
fixed $J$-value from that for the coherently rotating intermediate complex. Therefore,  energy of the coherent rotation as well as the associated moment of inertia and, thus, the characteristic length scale of the coherently rotating intermediate complex are also undefinable within the states with fixed $J$-value.

Let us assume that the coherent rotation may be considered
as the macroscopic motion. Then, an assumption of the spherical intermediate complex having rigid body moment of inertia is ruled out for, in this case,
$\hbar\omega\simeq 4$ MeV, resulting in the rotational energy ($\simeq 70$ Mev) which exceeds the total excitation energy ($\simeq 60-63$ MeV). Instead, the small value of $\hbar\omega\simeq 0.4$ MeV, corresponding to the period of the rotation $\simeq 10^{-20}$ sec, indicates an anomalously strong deformation of the coherently rotating complex with $J\simeq 34-38$. We calculate moment of inertia of this coherently rotating complex, ${\cal J}_{coh}$, and find that it corresponds to a chain-state of
the length $\simeq 30$ fm. This is close to the length  $\simeq 24$ fm of the chain of four touching carbon nuclei, calculated with $r_0=1.3$ fm.
We evaluate deformation energy of the chain as a sum of Coulomb energy of the two touched $^{24}Mg$ nuclei and double Coulomb energy of the two touched $^{12}C$ nuclei.  For $r_0$=1.3 fm, this gives $\simeq$38 MeV for the deformation energy. The energy of the coherent rotation, for $J\simeq 34-38$, is $\simeq 7$ MeV. Therefore, the energy of the intrinsic excitation (heat) is estimated $\geq 15$ MeV, {\sl i.e.} our ``carbon sausages" are really hot.

Let us evaluate average level spacing ($D$) of the intermediate complex with  $J\simeq 34-38$. We use the standard statistical model formula (the Weisskopf evaluation
\cite{Land37}, \cite{BlattWeiss52}, \cite{LandSmor55}) for the total decay width with fixed $(J,\pi)$-values: $\Gamma=(1/2\pi)Dn{\bar T}$. Here, $n$ is a number of the open channels and ${\bar T}$ is the averaged over the channels transmission coefficient. The total number of channels, specified by the channel spins, orbital momenta and  internal quantum numbers of the fragments, for the $^{24}Mg+^{24}Mg$ elastic and inelastic scattering, up to excitation of the
(6$^+$,4$^+$) states of the $^{24}Mg$ collision fragments, is $57$. We estimate number of the $\alpha$-transfer and two-$\alpha$-transfer channels and the proton and $\alpha$-emission
channels to be $\simeq 200$, where we have taken into account all possible the channel spins and orbital momenta for given internal quantum numbers of the fragments. Then our estimate is $n\simeq 250$.
Since the heavy ion collisions under consideration are characterized by a closeness to the potential barrier (centrifugal plus Coulomb barriers) or even are sub-barrier, and the proton and $\alpha$-particle evaporation is mostly sub-barrier,
we estimate ${\bar T}\simeq 0.5$. Then, for $\Gamma=0.2$ MeV, we obtain $D\simeq$10 keV. This indicates the relatively weak coupling to the continuum,
$\Gamma/(nD)\simeq 0.08$, in the regime of strongly overlapping resonances, $\Gamma /D\simeq 20$. Therefore the Bethe diagonal approximation (Section 56D
in \cite{Bethe37}) obtained from the Bethe unitary $S$-matrix for binary reactions involving complex collision partners is applicable (to be presented in the extended version of this Letter).

Note that the standard statistical model evaluation \cite{Bethe37}, \cite{LandSmor55}, \cite{BohrMott}, for the intrinsic excitation of 15 MeV of the intermediate nucleus, yields $D\simeq$0.01-0.1 keV. This is 2-3 orders of magnitude smaller than the average level spacing evaluated above. We interpret this reduction of $D$ as an indication
of a sizable admixture of clustering configurations in the wave function of the highly excited strongly deformed intermediate complex with large spin values $J\simeq 34-38$. These clustering configurations should not be considered as vibrational modes of the linear chain of four carbon nuclei each being in its ground state. Indeed it seems hardly probable that such a configuration would be stable with respect to decay. Therefore, it is perhaps more realistic to consider the hyper-deformed highly excited
states as the approximately linear chain of four relatively heavy clusters in combination with some ``valent" nucleons and $\alpha$-particles.

Up to now the idea of rotational coherence of ergodic, with respect to the energy and phase relaxation within the states with fixed $J$-values, nuclear molecules in continuum has been employed to extract spectroscopic information ($\hbar\omega,{\cal J}_{coh}$) from dynamics for a few colliding systems, ${\sl e.g.}$,
  $^{58}Ni+^{58}Ni$ \cite{Cindro},   $^{58}Ni+^{62}Ni$ \cite{Cindro}, \cite{ZPhysTiNi}, $^{46}Ti+^{58}Ni$ \cite{ZPhysTiNi},
$^{12}C+^{24}Mg$ \cite{PRCrapid99}, \cite{PRC01}, \cite{IJMPE03}, \cite{PRC06},
 $^{24}Mg+^{24}Mg$ and $^{28}Si+^{28}Si$ \cite{PRL99}, $^{24}Mg+^{28}Si$ \cite{PAN07}. Therefore, it may be of interest to apply this method for analysis of a large number of available data sets for many heavy ion colliding systems.
 On the other hand, the question is whether the stable coherent rotation survives violation of rotational symmetry and other external perturbations of the Hamiltonian?  The conjecture \cite{arXiv13} offers a possible specification of this question in terms of the correlation properties of eigenstates of the intermediate complex.

 Experimental manifestation of rotational coherent motion in complex molecules was initially met with scepticism ``because of the general belief
 that Coriolis interactions, anharmonicity and other interactions would destroy the coherence." (see subsection ``Changing a Dogma: Development of RCS" in
 \cite{ZewailNobLect}).
 The present discussion leads us to ask:
 Is rotational coherence of complex molecules necessarily destroyed in the conventionally statistical ergodic limit of
 structureless (non-selective) continuum under the conditions of complete intramolecular energy redistribution (thermalization) and vibrational dephasing in the regime of strong ro-vibrational coupling?
Though challenging experimentally, the subject is of interest to search for fingerprints of transformation of the quantum coherent wave packet dynamics
into macroscopic motion. More specifically, in the quantum regime, the spreading of the wave packets is determined by $\beta$-width, which has
essentially quantum origin of the spin off-diagonal correlations \cite{arXiv13}. Then what is a classical analog of $\beta$-width, for classically chaotic systems, when macroscopic non-linear dynamics takes over from the coherent rotation of the spreading quantum wave packets?
Addressing the questions above should include a comparison of the nature
of the dephasing analyzed in \cite{ZewailV1V2} with that discussed in this Letter (see also \cite{arXiv13} and references therein).

\section*{Acknowledgments}
We thank Sergei Maydanyuk for useful discussions.
Writing this Letter started when one of us (S.K.) was supported by Chinese Academy of Sciences Visiting Professorship for Senior International Scientists
(2012-2013).





\bibliographystyle{model1a-num-names}
\bibliography{<your-bib-database>}

\begin{thebibliography}{00}


\bibitem{ZewailV1V2}A. Zewail, Femtochemistry, Vols. I and II, World Scientific, Singapore, 1994.

\bibitem{ZewailNobLect}A. Zewail, www.nobelprize.org/nobel-prizes/chemistry/laureates/1999/zewail-lecture.pdf, and references therein.

\bibitem{CasChirI}G. Casati, B.V. Chirikov, Phys. Rev. Lett. 75 (1995) 350.

\bibitem{RMW10}G.E. Mitchell, A. Richter, H.A. Weidenm\"uller,
Rev. Mod. Phys. 82 (2010) 2845, and references therein.

\bibitem{arXiv13}S. Kun, Y. Li, H. Zhao, M.R. Huang, e-print arXiv: quant-ph/1307.4490.

\bibitem{PRCrapid99}S.Yu. Kun, A.V. Vagov, O.K. Vorov, Phys. Rev. C 59 (1999) R585.

\bibitem{PRC01}S.Yu. Kun, A.V. Vagov, W. Greiner, Phys. Rev. C 63 (2001) 014608.

\bibitem{IJMPE03}S.Yu. Kun, A.V. Vagov, L.T. Chadderton, W. Greiner,
Int. J. Mod. Phys. E 11 (2002) 273.

\bibitem{PRC06}L. Benet, S.Yu. Kun, Wang Qi, Phys. Rev. C 73 (2006) 064602, e-print arXiv: quant-ph/0503046.

\bibitem{PLB83}R.W. Zurm\"uhle, P. Kutt, R.R. Betts, S. Saini, F. Haas, Ole Hansen,
Phys. Lett. B 129 (1983) 384.

\bibitem{IndiaZPA}A. Sarma, R. Singh, Z. Phys. A 329 (1988) 195.

\bibitem{PRL99}S.Yu. Kun, B.A. Robson, A.V. Vagov, Phys. Rev. Lett. 83 (1999) 504.

\bibitem{ErMayer66}T. Ericson, T. Mayer-Kuckuk, Ann. Rev. Nucl. Sci. 16 (1966) 183.

\bibitem{DallHall66}P.G. Dallimore, I. Hall, Nucl. Phys. 88 (1966) 193.

\bibitem{PLB87}S. Saini, R.R. Betts, R.W. Zurm\"uhle, P.H. Kutt, B.K. Dichter, Phys. Lett. B 185 (1987) 316.

\bibitem{ZPhysII}S.Yu. Kun, Z. Phys. A 357 (1997) 271.

\bibitem{Bethe37}H.A. Bethe, Rev. Mod. Phys. 9 (1937) 69.

\bibitem{Koonin89}S.E. Koonin, W. Bauer, A. Sch\"afer, Phys. Rev. Lett. 62 (1989) 1247.

\bibitem{FKK80}H. Feshbach, A. Kerman, S.E. Koonin, Ann. Phys. 125 (1980) 429.

\bibitem{Str57}V.M. Strutinsky, in Proceeding of Conference on Nuclear Reactions at Low and Medium Energies,
Moscow 1957, p. 522, Academy of Sciences USSR Publishing, 1958.

\bibitem{ErStr58plus}T. Ericson, V.M. Strutinsky, Nucl. Phys. 8 (1958) 284; 9 (1958/1959) 689.

\bibitem{ZPhysI}S.Yu. Kun, Z. Phys. A 357 (1997) 255.

\bibitem{HellerPRL84}E.J. Heller,  Phys. Rev. Lett. 53 (1984) 1515.

\bibitem{PRC90}A.H. Wuosmaa, R.W. Zurm\"uhle, P. Kutt, S.F. Pate,
 S. Saini, M.L. Halbert, D.C. Hensley,
Phys. Rev. C 41 (1990) 2666.

\bibitem{PRA07}L. Benet, L.T. Chadderton, S.Yu. Kun, Wang Qi, Phys. Rev. A 75 (2007) 062110; e-print arXiv: quant-ph/0610091.

\bibitem{Feynman65} R. Feynman, The Character of Physical Law, based on the Messenger Lectures given by R. Feynman at Cornell University
in 1964, Lecture 6: Probability and Uncertainty - the Quantum Mechanical view of Nature, first published by BBC, 1965.

\bibitem{JPhChemA06} P.D.D. Monks, J.N.L. Connor, S.C. Althorpe, J. Phys. Chem. A 110 (2006) 741.

\bibitem{JPhChemA07} P.D.D. Monks, J.N.L. Connor, S.C. Althorpe, J. Phys. Chem. A 111 (2007) 10302.

\bibitem{Land37}L.D. Landau, Phys. Zeits. Sow. 11 (1937) 556.

\bibitem{BlattWeiss52}J.M. Blatt, V.F. Weisskopf, Theoretical Nuclear Physics, Wiley, 1952.

\bibitem{LandSmor55}L. Landau, Ya. Smorodinsky, Lectures on Theory of Atomic Nucleus, Moscow, Technical-Theoretical Publishing, 1955.

\bibitem{BohrMott}A. Bohr, B. Mottelson, Nuclear Structure, Vol. 1, Benjamin, New York, 1969.

\bibitem{Cindro}L. Vannucci,  U. Abbondanno, M. Bettiolo, M. Bruno, N. Cindro, M. D'Agostino, P.M. Milazzo, R.A.
Ricci, T. Ritz, W. Scheid, G. Vannini, Z. Phys. A 355 (1996) 41.

\bibitem{ZPhysTiNi}S.Yu. Kun, U. Abbondanno, M. Bruno, N. Cindro, M. D'Agostino, P.M. Milazzo, R.A.
Ricci, T. Ritz, B.A. Robson, W. Scheid, A.V. Vagov, G. Vannini, L. Vannucci,
Z. Phys. A 359 (1997) 145.

\bibitem{PAN07}L. Benet, L.T. Chadderton, S.Yu. Kun, O.K. Vorov, Wang Qi,
Phys. At. Nucl. 71 (2008) 819, e-print arXiv: 0705.0502.

\end{thebibliography}



\end{document}